\documentclass[%
 reprint,
 amsmath,amssymb,
 aps,
]{revtex4-2}
\usepackage[utf8]{inputenc}
\usepackage{graphicx}
\usepackage{dcolumn}
\usepackage{bm}
\usepackage{physics}
\usepackage{url}
\usepackage{hyperref}
\hypersetup{
 setpagesize=false,
 bookmarksnumbered=true,%
 bookmarksopen=true,%
 colorlinks=true,%
 linkcolor=blue,
 citecolor=blue,
}

\begin{document}

\preprint{APS/123-QED}

\title{Charge-dependent anisotropic flow in high-energy heavy-ion collisions from relativistic resistive magneto-hydrodynamic expansion}

\author{Kouki Nakamura$^{1,2}$}
 \email{knakamura@hken.phys.nagoya-u.ac.jp}
\author{Takahiro Miyoshi$^{2}$}%
 \email{miyoshi@sci.hiroshima-u.ac.jp}
\author{Chiho Nonaka$^{1,2,3}$}%
 \email{nchiho@hiroshima-u.ac.jp}
\author{Hiroyuki R. Takahashi$^{4}$}%
 \email{takhshhr@komazawa-u.ac.jp}
\affiliation{%
$^1$Department of Physics, Nagoya University, Nagoya 464-8602, Japan \\
$^2$Department of Physics, Hiroshima University, Higashihiroshima 739-8526, Japan\\
$^3$Kobayashi Maskawa Institute, Nagoya University, Nagoya 464-8602, Japan\\
$^4$Department of Physics, Komazawa University, Tokyo 154-8525, Japan\\
 }%




\date{\today}

\begin{abstract}
We have investigated the charge-dependent anisotropic flow in high-energy heavy-ion collisions, using relativistic resistive magneto-hydrodynamics (RRMHD).
We consider the optical Glauber model as an initial model of the quark-gluon plasma (QGP) and the solution of the Maxwell equations with source term of the charged particles in two colliding nuclei as initial electromagnetic fields.
The RRMHD simulation is performed with these initial conditions in Au-Au and Cu-Au collisions at $\sqrt{s_{\mathrm{NN}}} = 200$ GeV.
We have calculated the charge-odd contribution to the directed flow $\Delta v_1$ and elliptic flow $\Delta v_2$ in both collisions based on electric charge distributions as a consequence of RRMHD.
Our results show that the $\Delta v_1$ and $\Delta v_2$ are approximately proportional to the electrical conductivity ($\sigma$) of the medium.
In the $\sigma=0.023~\mathrm{fm}^{-1}$ case, our result of $\Delta v_1$ is consistent with STAR data in Au-Au collisions.
Furthermore, in Cu-Au collisions, $\Delta v_1$ has a non-zero value at $\eta = 0$.
We conclude that the charge-dependent anisotropic flow is a good probe to extract the electrical conductivity of the QGP medium in high-energy heavy-ion experiments.

\end{abstract}

\maketitle
\section{introduction}
In high-energy heavy-ion collisions, the production of ultraintense electromagnetic fields by two colliding nuclei is one of the hottest topics~\cite{PhysRevC.85.044907, PhysRevC.88.024911, BZDAK2012171}.
For example, in $\sqrt{s_{NN}} = 200$ GeV Au-Au collisions at Relativistic Heavy Ion Collider (RHIC), the highest intensity of the magnetic field in our universe may be reached, e.g., $|eB|\sim 10^{15} ~\mathrm{T}$~\cite{KHARZEEV2008227, doi:10.1142/S0217751X09047570, PhysRevC.83.054911, MCLERRAN2014184, Huang_2016}.
The intensity of the magnetic field in the transverse plane increases approximately linearly with the center of mass collision energy~\cite{PhysRevC.85.044907, PhysRevC.88.024911, BZDAK2012171}. 
The corresponding electric field in the transverse plane is also enhanced by a Lorentz factor of colliding nuclei.
Such intense electromagnetic fields can affect the hadron distribution detected in high-energy heavy-ion collision experiments at RHIC and the Large Hadron Collider (LHC).
As the electromagnetic response of the quark-gluon plasma (QGP), electromagnetic fields affect the electric charge of quarks.
As a consequence of it, the charge dependence is found in directed flows of hadrons at RHIC and the LHC~\cite{STAR:2008jgm, STAR:2017ykf, PhysRevLett.125.022301}.
Furthermore, the presence of strong electromagnetic fields leads to novel quantum phenomena such as chiral magnetic effect (CME)~\cite{PhysRevD.78.074033} and chiral magnetic wave (CMW)~\cite{PhysRevD.83.085007}.
There are several efforts to detect these phenomena in the iso-bar experiments such as Zr-Zr and Ru-Ru collisions at $\sqrt{s_{\mathrm{NN}}}=200$ GeV~\cite{PhysRevC.105.014901}.
Also, the effect of the background magnetic field on the quantum chromodynamics (QCD) phase diagram is investigated~\cite{MCINNES2016173}.

For the description of time evolution of the initial electromagnetic fields in dynamics of high-energy heavy-ion collisions, construction of the relativistic resistive magneto-hydrodynamics (RRMHD) is indispensable~\cite{Nakamura:2022idq, Nakamura:2022wqr}.
The RRMHD framework describes the dynamics of the plasma with finite electrical conductivity coupled with electromagnetic fields.
In the RRMHD framework, Ohm's law is considered to close the system of differential equations.
One of the possibilities for building a model based on RRMHD is to employ an ideal limit of Ohm's law which assumes the infinite electrical conductivity of the medium, as well known relativistic ideal MHD.
In the analysis based on relativistic ideal MHD~\cite{Inghirami:2016iru, Inghirami:2019mkc}, the effect of the magnetic field has only a very small impact on the collective flow of charged hadrons.
In our previous study~\cite{Nakamura:2022idq}, in analysis based on RRMHD with finite electrical conductivity, we show the sizable effect of the dissipation associated with Ohm's law on the directed flow of hadrons in asymmetric collision systems such as Cu-Au collisions. 
We have found that the conduction current induced by Ohm's law plays an important role in the dynamics of high-energy heavy-ion collisions. 

The electrical conductivity of the QCD matter is estimated to be, $\sigma = (5.8\pm2.9)/\hbar c$ fm$^{-1}$ in three flavor QGP at temperature $T=250$ MeV, by the lattice QCD calculation which is the first principle simulation of QCD~\cite{Aarts:2007wj, Ding:2010ga, Brandt:2012jc, FRANCIS2012212}.
It indicates a possibility of the long-time electromagnetic response of the QGP medium. 
For example, at RHIC, the time scale of the dissipation associated with Ohm's law, $\tau_\sigma\sim 1/\sigma\sim 10^2$ fm is sufficiently larger than the time scale of the dynamics of high-energy heavy-ion collisions, $\tau_f\sim 10$ fm.
Namely, we should take into account the finite electrical conductivity in RRMHD.
Also, the incomplete electromagnetic response of QCD matter is discussed by introducing the relaxation time of the electric current in Refs.~\cite{Akamatsu:2011nr, PhysRevC.105.L041901, Dash:2022xkz}.
It suggests that the electric current is suppressed by the long-time electromagnetic response associated with the relaxation process of the electric current. 

In this paper, we investigate the effect of electromagnetic fields on the charge-dependent anisotropic flow based on RRMHD.
We apply the newly developed RRMHD simulation code to high-energy heavy-ion collisions~\cite{Nakamura:2022idq, Nakamura:2022wqr}.
We employ the optical Glauber model~\cite{Glauber} as an initial condition of the QGP medium.
The solution of Maxwell equations with the source term of the charged particles inside of the two colliding nuclei is taken to be the initial electromagnetic fields~\cite{PhysRevC.88.024911}.
We shall discuss the electrical conductivity dependence of the charge-dependent anisotropic flow.

This paper is organized as follows.
In Sec.~\ref{RRMHD}, we show the RRMHD model for high-energy heavy-ion collisions.
Numerical results are shown in Sec.~\ref{results} and a summary is given at the end in Sec.~\ref{summary}.
Unless otherwise specified, we use natural units $\hbar = c = \epsilon_0 = \mu_0 = 1$, where $\epsilon_0$ and $\mu_0$ are the electric permittivity and the magnetic permeability in a vacuum, respectively.
Throughout the paper, the components of the four tensors are indicated with greek indices, whereas three vectors are denoted as boldface symbols. 

\section{Relativistic resistive magneto-hydrodynamic model}\label{RRMHD}

\subsection{Relativistic resistive magnetohydrodynamic equation}
The RRMHD equation consists of the conservation laws for the charged current $N^\mu$ and for the total energy-momentum tensor of the plasma $T^{\mu\nu}$ in the dynamics of the whole system.
They are written by,
\begin{eqnarray}
  \label{ccc}
  \nabla_\mu N^\mu = 0,\\
  \label{emc}
  \nabla_\mu T^{\mu\nu} = 0,
\end{eqnarray}
where $\nabla_\mu$ is the covariant derivative.
The electromagnetic fields satisfy Maxwell equations,
\begin{eqnarray}
  \label{maxwell1}
  \nabla_\mu F^{\mu\nu} = -J^{\nu},\\
  \label{maxwell2}
  \nabla_\mu ~^\star F^{\mu\nu} = 0,
\end{eqnarray}
where $F^{\mu\nu}$ is a Faraday tensor and $^{\star}F^{\mu\nu} = \frac{1}{2}\epsilon^{\mu\nu\rho\sigma}F_{\rho\sigma}$ is its dual tensor,
with $\epsilon^{\mu\nu\rho\sigma} = (-g)^{-1/2}[\mu\nu\rho\sigma]$, $g = \det(g_{\mu\nu}) $ and $[\mu\nu\rho\sigma]$ is a completely anti-symmetric tensor. 
Here $g_{\mu\nu}$ is a metric tensor and we take the metric $g^{\mu\nu} = \rm{diag}(-1,1,1,1)$ in the Minkowski space-time.

To close the system of equations~Eqs.~(\ref{ccc})-(\ref{maxwell2}), we employ the simplest form of Ohm’s law in Ref.~\cite{Blackman:1993pbp}.
In the covariant form, Ohm's law is written by,
\begin{equation}\label{ohmslaw}
  J^\mu = \sigma F^{\mu\nu}u^{\nu} + qu^\mu,
\end{equation}
where $\sigma$ is electrical conductivity, $u^\mu$ is the four-velocity of the fluid, and $q = -J^\mu u_\mu$ is the electric charge density of the fluid in the comoving frame.
Maxwell equations lead to the charge conservation law,
\begin{equation}
    \partial_\mu J^\mu = 0.
\end{equation}

We numerically solve the system of RRMHD equations~Eqs.~(\ref{ccc})-(\ref{maxwell2}) in the Milne coordinates $(\tau,\bm{x}_{\mathrm{T}},\eta_s)$.
We have introduced variables; $\tau = \sqrt{t^2 - z^2}$ is a longitudinal proper time, $\bm{x}_{\mathrm{T}} = (x,y)$ represents transverse coordinates, and $\eta_s = \frac{1}{2}\ln\left(\frac{t+z}{t-z}\right)$ is a space rapidity.
As an equation of state (EoS), we employ the ideal gas EoS, $p = e/3$, for simplicity.
We note that we neglect the effect of the electric charge density in the EoS since the electric charge density is much smaller than the energy density of the fluid.
In the previous study~\cite{Nakamura:2022idq,Nakamura:2022wqr}, we constructed RRMHD model for high-energy heavy-ion collisions.
Our RRMHD model is adopted for the estimation of the charge-dependent anisotropic flow in high-energy heavy-ion collisions.

\subsection{Initial model}
\subsubsection{Medium}

We consider the optical Glauber model~\cite{Glauber} as an initial condition of the QGP medium.
In the optical Glauber model, the energy density takes the form,
\begin{equation}
   e(\bm{x}_\perp,\eta_s) = e_0M(\bm{x}_\perp)f_{\rm{tilt}}(\eta_s),
\end{equation}
where $e_0= 55~\mathrm{GeV/fm}^3$\cite{Inghirami:2019mkc} is the energy density at $(\bm{x}_{\perp},\eta_s) = (\bm{0}~\mathrm{fm},0)$ and $f_{\mathrm{tilt}}(\eta_s)$ is a longitudinal profile function with the tilted sources~\cite{Bozek:2010bi}. 
For a tilted initial energy density distribution~\cite{Bozek:2010bi}, we have introduced the function $M(\bm{x}_\perp;\bm{b})$ as below,
\begin{equation}
  M(\bm{x}_\perp,\eta_s;\bm{b}) = \frac{(1-\alpha_{\rm{H}})W_N(\bm{x}_\perp,\eta_s;\bm{b}) + \alpha_{\rm{H}}n_{\rm{coll}}(\bm{x}_\perp;\bm{b})}{(1-\alpha_{\rm{H}})W_N(\bm{0},0;\bm{0}) + \alpha_{\rm{H}}n_{\rm{coll}}(\bm{0};\bm{0})},
\end{equation}
where $\alpha_\mathrm{H} = 0.05$~\cite{Inghirami:2019mkc} is a collision hardness parameter and $n_{\mathrm{coll}}$ is the number of binary nucleon collisions.
We have defined the wounded nucleon's weight function $W_N$ as,
\begin{equation}
  W_N(\bm{x}_\perp,\eta_s;\bm{b}) = 2(n^A_{\rm{part}}(\bm{x}_\perp;\bm{b})f_-(\eta_s) + n^B_{\rm{part}}(\bm{x}_\perp;\bm{b})f_+(\eta_s)),
\end{equation} 
where,
\begin{equation}
  f_-(\eta_s) =
  \begin{cases}
    1 & \text{($\eta_s < -\eta_m $)} \\
    \frac{-\eta_s + \eta_m}{2\eta_m} & \text{($-\eta_m\leq \eta_s \leq \eta_m $),} \\
    0 & \text{($\eta_s > \eta_m$)}
  \end{cases}
\end{equation}
and,
\begin{equation}
  f_+(\eta_s) =
  \begin{cases}
    0 & \text{($\eta_s < -\eta_m $)} \\
    \frac{\eta_s + \eta_m}{2\eta_m} & \text{($-\eta_m\leq \eta_s \leq \eta_m $),} \\
    1 & \text{($\eta_s > \eta_m$)}
  \end{cases}
\end{equation}
where $\eta_m = 3.36$~\cite{Bozek:2010bi} is a parameter.
We define the tilted longitudinal profile function $f_{\rm{tilt}}(\eta_s)$ as,
\begin{equation}
  f_{\rm{tilt}}(\eta_s) = \exp\left(\frac{-(|\eta_s| - \eta_{\rm{flat}}/2)^2}{2w_\eta^2}\theta(|\eta_s| - \eta_{\rm{flat}}/2)\right),
\end{equation}
where $w_{\eta} = 4.0$~\cite{Inghirami:2019mkc} is  a width of the gauss function in $f_\mathrm{tilt}(\eta_s)$ and $\eta_{\mathrm{\rm{flat}}} = 5.9$~\cite{Inghirami:2019mkc} is a width of the plateau for the rapidity distribution.
The parameters of the initial condition of the QGP medium have been determined from the comparison with the STAR data of the directed flow in Au-Au collisions~\cite{STAR:2008jgm, Nakamura:2022idq}.
The values of these parameters are consistent with those of Refs.~\cite{Bozek:2010bi,Inghirami:2019mkc}.
The energy density profiles in Au-Au and Cu-Au collisions are shown in Figs.~1 and 2 of our previous study~\cite{Nakamura:2022idq}.

\subsubsection{Electromagnetic field}
We take the solution of the Maxwell equations as the initial condition of electromagnetic fields~\cite{PhysRevC.88.024911}.
We consider the system in which the electric charge $q_0$ is moving along parallel to the beam axis ($\hat{\mathbf{z}}$) with velocity $v$ in the laboratory frame by an observer located at $\bm{r} = z\hat{\bm{z}} + \bm{x}_\perp$ in the Minkowski coordinates.
In such a system, the Maxwell equations are written by,
\begin{gather}
 \nabla\cdot \bm{B} = 0,~~~\nabla\times \bm{E} = -\frac{\partial \bm{B}}{\partial t},\\
\nabla\cdot \bm{D} = q_0\delta(z-vt)\delta(\bm{b}),\\
\nabla\times \bm{H} = \frac{\partial \bm{D}}{\partial t} + \sigma_0 \bm{E} + q_0v\hat{\bm{z}}\delta(z-vt)\delta(\bm{b}),
\end{gather}
where $\bm{B}$ is the magnetic field, $\bm{E}$ is the electric field, $\bm{H} = \mu\bm{B}$, and $\bm{D} = \epsilon\bm{E}$.
We take a constant permittivity $\epsilon = 1$, a constant permeability $\mu = 1$ and a constant finite electrical conductivity $\sigma_0 = 5.8~\mathrm{MeV}$~\cite{Ding:2010ga,Aarts:2007wj}.
If we set parameters $\gamma$, $\sigma_0$ and $b$ to be satisfied with the condition, $\gamma_0\sigma_0 b \gg 1$, the solutions of the Maxwell equations are written by,
\begin{eqnarray}
  E_r = B_\phi = \frac{q_0(\hbar c)^{3/2}}{2\pi}\frac{b\sigma_0/(\hbar c)}{4x_\pm^2}\exp\left(-\frac{b^2\sigma_0/(\hbar c)}{4x_\pm}\right)\nonumber,\\
  E_z = -\frac{q_0(\hbar c)^{3/2}}{4\pi}\frac{x_\pm - b^2\sigma_0/(4\hbar c)}{\gamma_0^2x_\pm^3}\exp\left(-\frac{b^2\sigma_0/(\hbar c )}{4x_\pm}\right),\nonumber\\
\end{eqnarray}
where we have introduced $\gamma_0 = 1/\sqrt{1 - v^2}$ and $x_{\pm} = t\pm v/z$.
To clarify the dimension of electromagnetic fields, $\mathrm{GeV^{1/2}/fm^{3/2}}$, we explicitly write $\hbar$ and $c$.
We assume the electric charge distribution inside two colliding nuclei as being uniform and spherical for simplicity.
The total electromagnetic fields are derived by integration over the inside of two colliding nuclei at each point of our computational grid.
The initial electromagnetic fields in Au-Au and Cu-Au collisions are discussed in Figs.~3 and 4 of our previous study~\cite{Nakamura:2022idq}.

\section{Numerical results}\label{results}
We perform the RRMHD simulation~\cite{Nakamura:2022idq, Nakamura:2022wqr} with these initial conditions in Au-Au and Cu-Au collisions.
We focus on the RHIC energy to compare our results with STAR data.
We start the RRMHD simulation at the proper time $\tau_0 = 0.4~\mathrm{fm}$ which is determined from the comparison with the STAR data of the directed flow in Au-Au collisions~\cite{Nakamura:2022idq}.
Since the directed flow has no electrical conductivity dependence~\cite{Nakamura:2022idq}, the $\tau_0$ has the same value in the cases of the $\sigma = 0.0058, 0.023,$ and $0.1~\mathrm{fm}^{-1}$. 
The value of $\tau_0$ is consistent with that of Ref.~\cite{Bozek:2010bi}.
We terminate the RRMHD simulation when the energy density of all fluid elements becomes below the freezeout energy density $e(\eta_s, \mathbf{x}_{\mathrm{T}}) = 0.15~\mathrm{GeV/fm}^3$.

\subsection{Charge distributions on the freezeout hypersurface}\label{charge distribution on the freezeout surface}
\begin{figure}[ht]
\includegraphics[width=8.5cm,height=6cm]{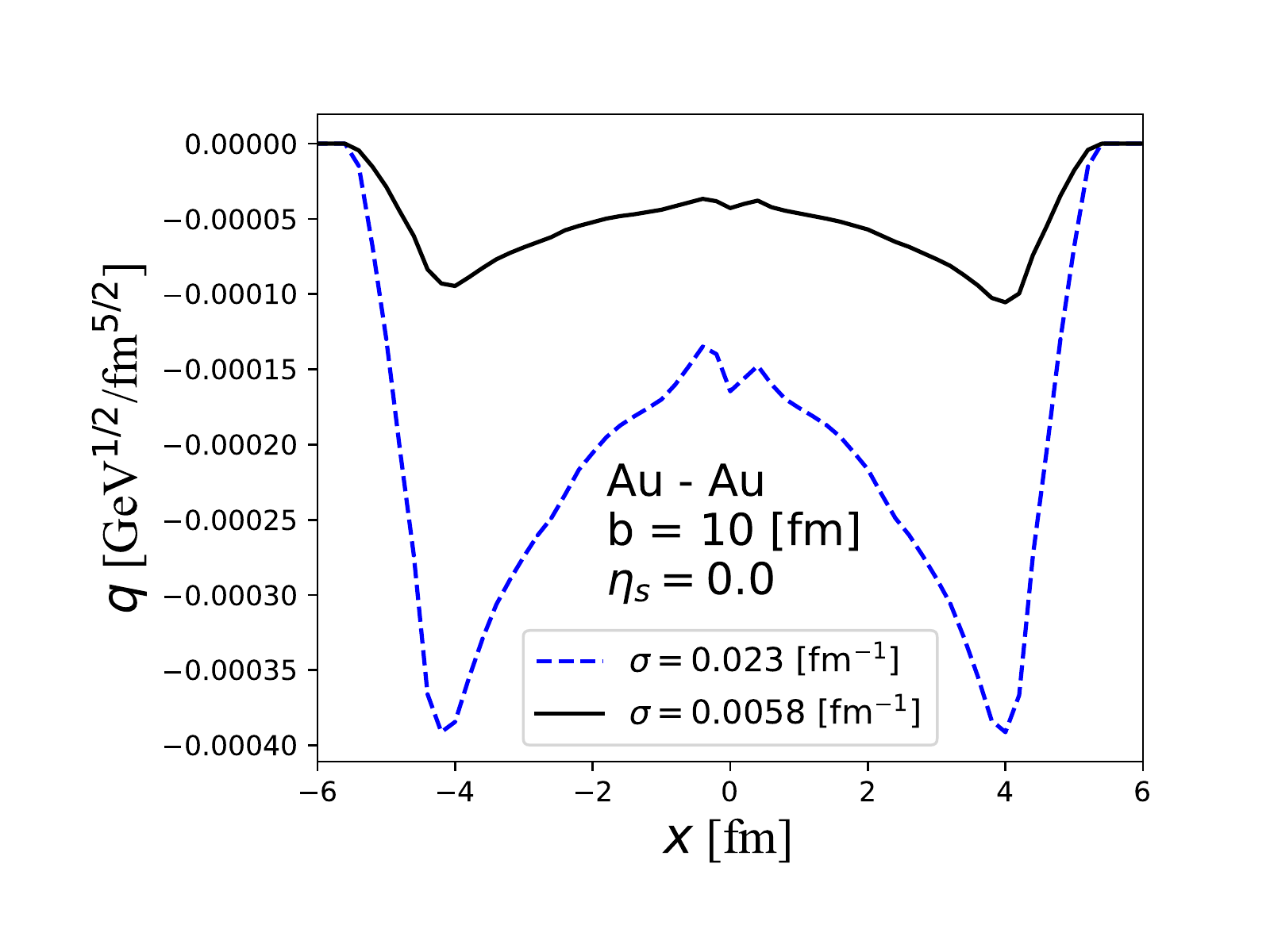}
\caption{\label{fig_hydro:q_freezeout surface eta0} (color online) The electric charge distributions on the freezeout hypersurface as a function of $x$ at $\eta_s = 0.0$ and $y = 0.0$ fm in Au-Au collisions. The black solid and blue dashed lines represent the charge density in the cases of $\sigma = 0.0058$ and $0.023~\mathrm{fm}^{-1}$. }
\end{figure}
\begin{figure*}[ht]
  \begin{minipage}[l]{0.45\linewidth}
    \includegraphics[width=8cm,height=6cm]{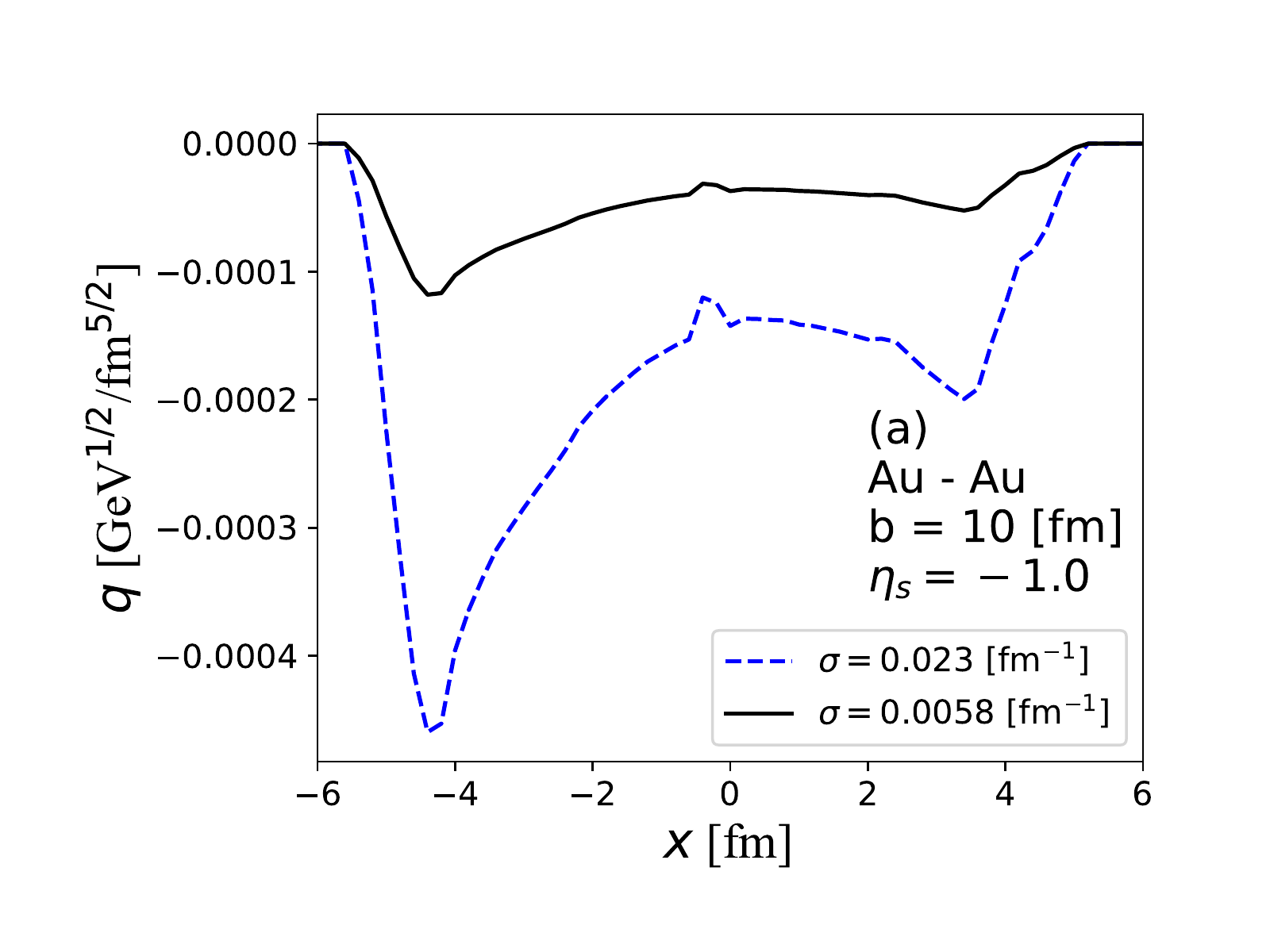}
  \end{minipage}
  \begin{minipage}[r]{0.45\linewidth}
    \includegraphics[width=8cm,height=6cm]{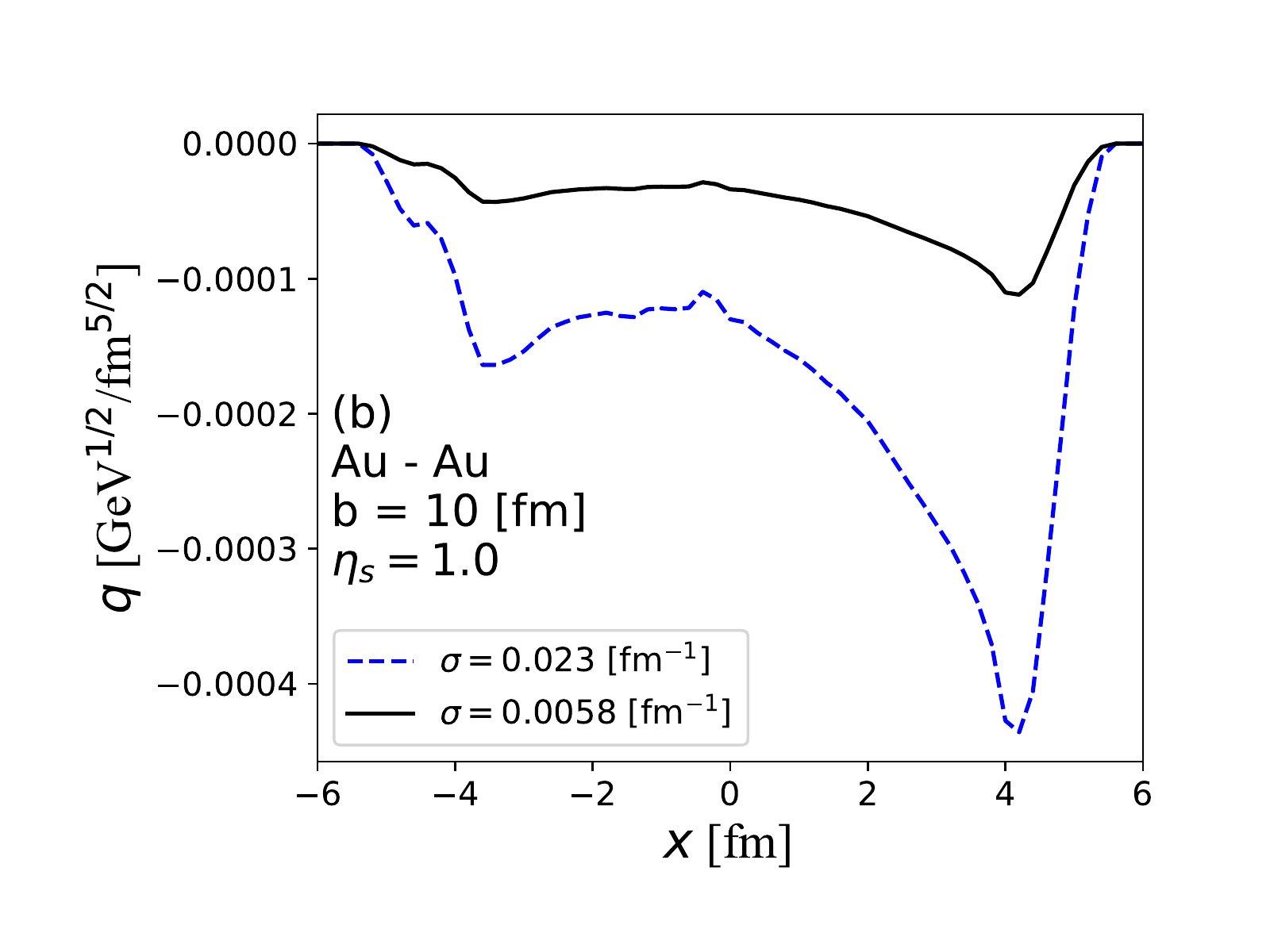}
     \end{minipage}
    \caption{(color online) The electric charge distributions on the freezeout hypersurface as a function of $x$ at $y = 0.0$ fm in the cases of (a) $\eta_s = -1.0$ and (b) $\eta_s = 1.0$ in Au-Au collisions. The black solid and blue dashed lines represent $\sigma = 0.0058$ and $0.023~ \mathrm{fm}^{-1}$ cases.
    }
    \label{fig_hydro:q_freezeout surface eta10}
\end{figure*}
Figure~\ref{fig_hydro:q_freezeout surface eta0} shows the electric charge distribution on the freezeout hypersurface at $(y,\eta_s) = (0~\mathrm{fm},0)$ in Au-Au collisions.
The freezeout hypersurface contains the information of velocity and location of the fluid elements of whole time steps at freezeout process where the hydrodynamic picture is switched to the particle picture.
Since the fluid elements at small $|x|$ have larger energy density than that at large $|x|$, they freeze out at a later time than that at large $|x|$.
Then, the electric charge density at small $|x|$ represents the electric charge density at a later time than that at large $|x|$.  
The centers of the colliding Au nuclei are located at $(x,y) = (\pm 5~\mathrm{fm}, 0~\mathrm{fm})$ in the transverse plane.
The Au nucleus located at $x = 5$ fm ($x = -5$ fm) moves to the forward (backward) space rapidity.
We will show the charge-dependent anisotropic flow in the cases of $\sigma = 0.0058,~0.023,$ and $0.1~\mathrm{fm}^{-1}$ in Sec.~\ref{Section: Charge-dependent anisotropic flow}.
However, here, we compare the electric charge distribution in $\sigma = 0.023~\mathrm{fm}^{-1}$ case with only that in $\sigma = 0.0058~\mathrm{fm}^{-1}$ case since the electric charge distribution in $\sigma = 0.1~\mathrm{fm}^{-1}$ case is qualitatively same as that in $\sigma = 0.0058~\mathrm{fm}^{-1}$ and $\sigma = 0.023~\mathrm{fm}^{-1}$ cases. 
The black and blue dashed lines represent the results in the cases of $\sigma = 0.0058$ and $0.023~\mathrm{fm}^{-1}$, respectively.
In Fig.~\ref{fig_hydro:q_freezeout surface eta0}, the negative charges are induced inside the freezeout hypersurface, since the electric field is facing outside the freezeout hypersurface.
Also, the negative charge distribution has two local minimums at $x\sim 4$ fm and $-4$ fm in both of electrical conductivity cases.
The location of the local minimums of the negative electric charge distribution is correlated with the shorter axis of the initial almond shape of the energy density of fluid in Fig.~1 (a) of our previous study~\cite{Nakamura:2022idq}.
Consequently, the momentum of the negatively charged hadrons on the freezeout hypersurface is correlated with the elliptic momentum anisotropy of the fluid induced by the almond-shaped pressure gradient on the freezeout hypersurface. 
On the other hand, the production of positively charged hadrons is suppressed due to the negative chemical potential of electric charge on the freezeout hypersurface. 
Then, this structure of the electric charge distribution may enhance the elliptic flow of negatively charged hadrons and reduce that of positively charged hadrons.

Figures~\ref{fig_hydro:q_freezeout surface eta10} (a) and (b) display the electric charge distribution on the freezeout hypersurface at $(y,\eta_s) = (0~\mathrm{fm},-1)$ and $(0~\mathrm{fm}, +1)$ in Au-Au collisions, respectively.
As shown in Fig.~4 (b) of Ref.~\cite{Nakamura:2022idq}, in $\eta_s < 0$, the electric field produced by the nucleus located at the negative $x$ side is dominant.
As a result, in Fig.~\ref{fig_hydro:q_freezeout surface eta10} (a), the negative charge density at the negative $x$ side is larger than that at the positive $x$ side.
The electric charge density has the largest local minimum at $x = -4.0$ fm.
On the other hand, as shown in Fig.~4 (b) of Ref.~\cite{Nakamura:2022idq}, in $\eta_s > 0$, the electric field produced by the nucleus located at the positive $x$ side has a strong influence on electric charge distributions.
The electrical conductivity dependence is clearly observed in all space rapidity regions.
Because the absolute value of electric current is proportional to electrical conductivity, the electric charge density with higher electrical conductivity becomes larger.
The induced electric charge density is approximately proportional to the electrical conductivity.
It is consistent with Ohm's law.

\begin{figure}[ht]
\includegraphics[width=8.5cm,height=6cm]{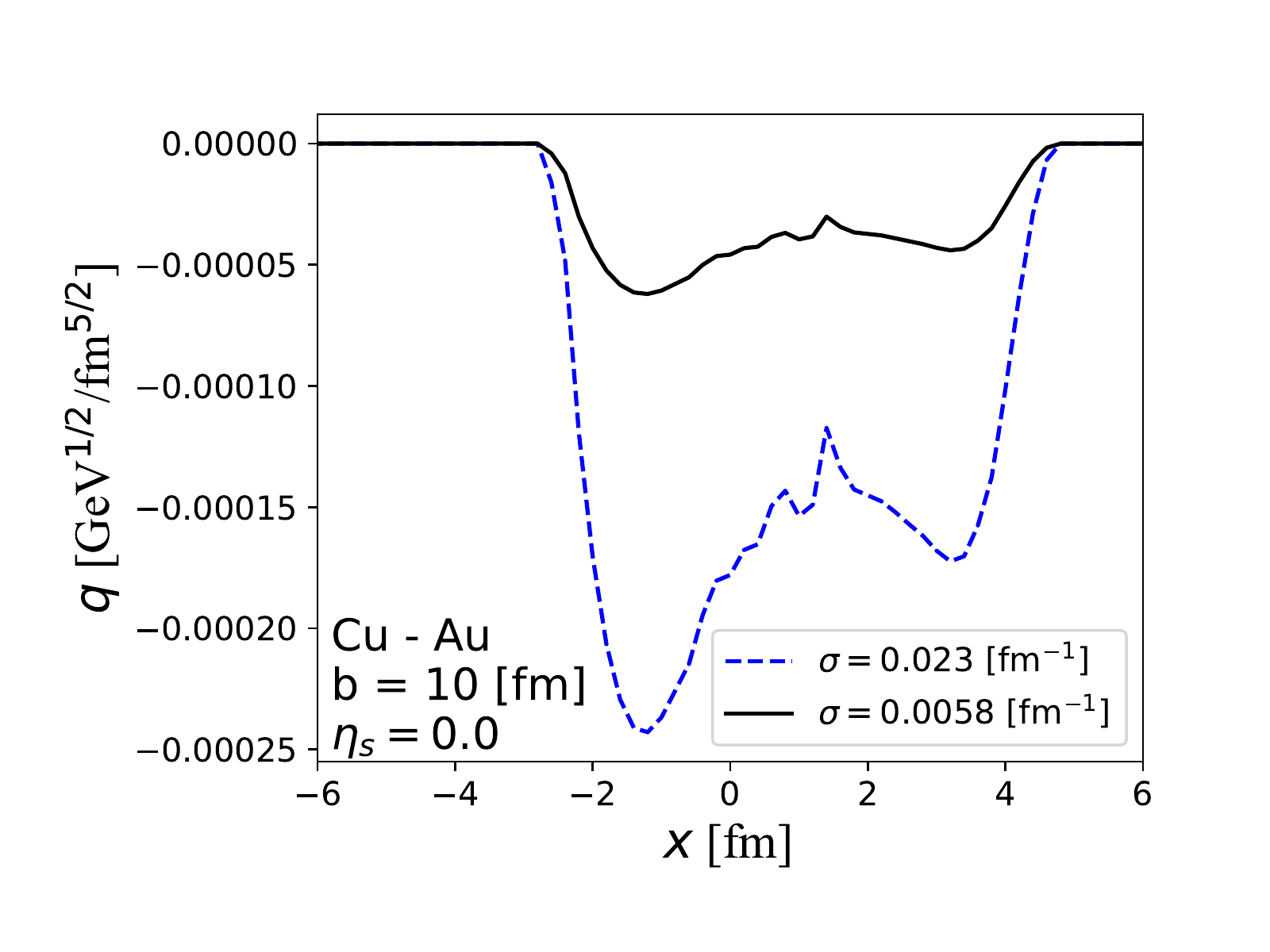}
\caption{\label{fig_hydro:q_freezeout surface eta0 Cu-Au} (color online) The electric charge distributions on the freezeout hypersurface as a function of $x$ at $\eta_s = 0.0$ and $y = 0.0$ fm in Cu-Au collisions. The black solid and blue dashed lines represent the charge distributions in the cases of $\sigma = 0.0058$ and $0.023~\mathrm{fm}^{-1}$. }
\end{figure}
Figure~\ref{fig_hydro:q_freezeout surface eta0 Cu-Au} represents the electric charge distribution on the freezeout hypersurface at $(y,\eta_s) = (0~\mathrm{fm},0)$ in Cu-Au collisions.
The centers of Cu and Au nucleus in the transverse plane are located at $(x,y) = (+ 5~\mathrm{fm},0~\mathrm{fm})$ and $(-5~\mathrm{fm},0)$, respectively.
We define that the Cu (Au)-going side is forward (backward) rapidity.
Because of the asymmetry of the electric field in Cu-Au collisions, the electric charge distribution has an asymmetric structure at $\eta_s = 0$.
As shown in Fig.~6 (b) of Ref.~\cite{Nakamura:2022idq}, in the transverse plane, the non-zero electric field facing the center of Cu nucleus in the freezeout hypersurface induces the negative electric charge in the negative $x$ region. 
Thus, the negative charge density at the negative $x$ region has a larger value than that at the positive $x$ region.
This is different from that in Au-Au collisions in Fig.~\ref{fig_hydro:q_freezeout surface eta0}.
It induces a non-zero value of the charge-odd contribution to the directed flow at $\eta_s=0$. 

\begin{figure*}[t]
  \begin{minipage}[l]{0.45\linewidth}
    \includegraphics[width=8cm,height=6cm]{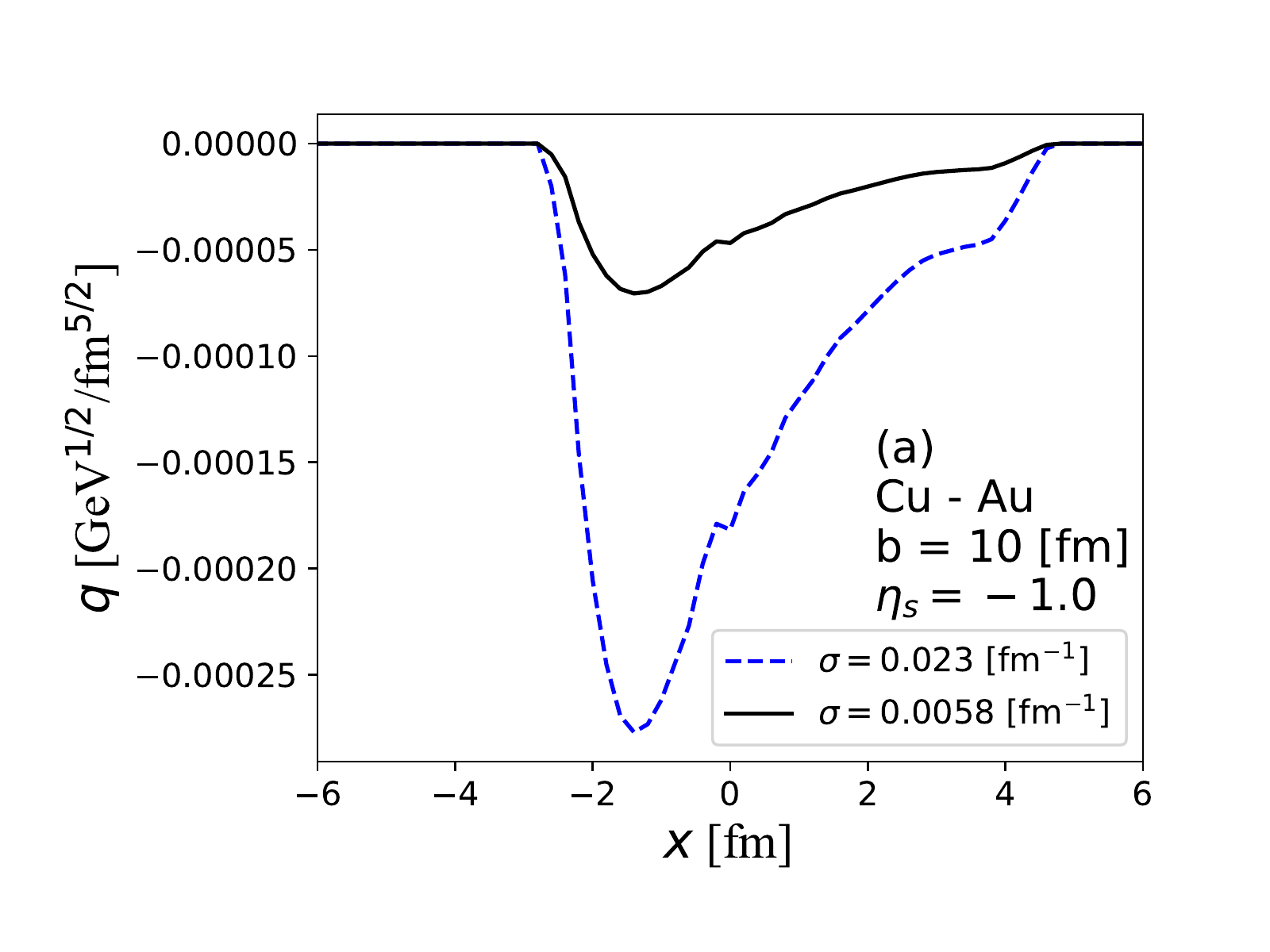}
  \end{minipage}
  \begin{minipage}[r]{0.45\linewidth}
    \includegraphics[width=8cm,height=6cm]{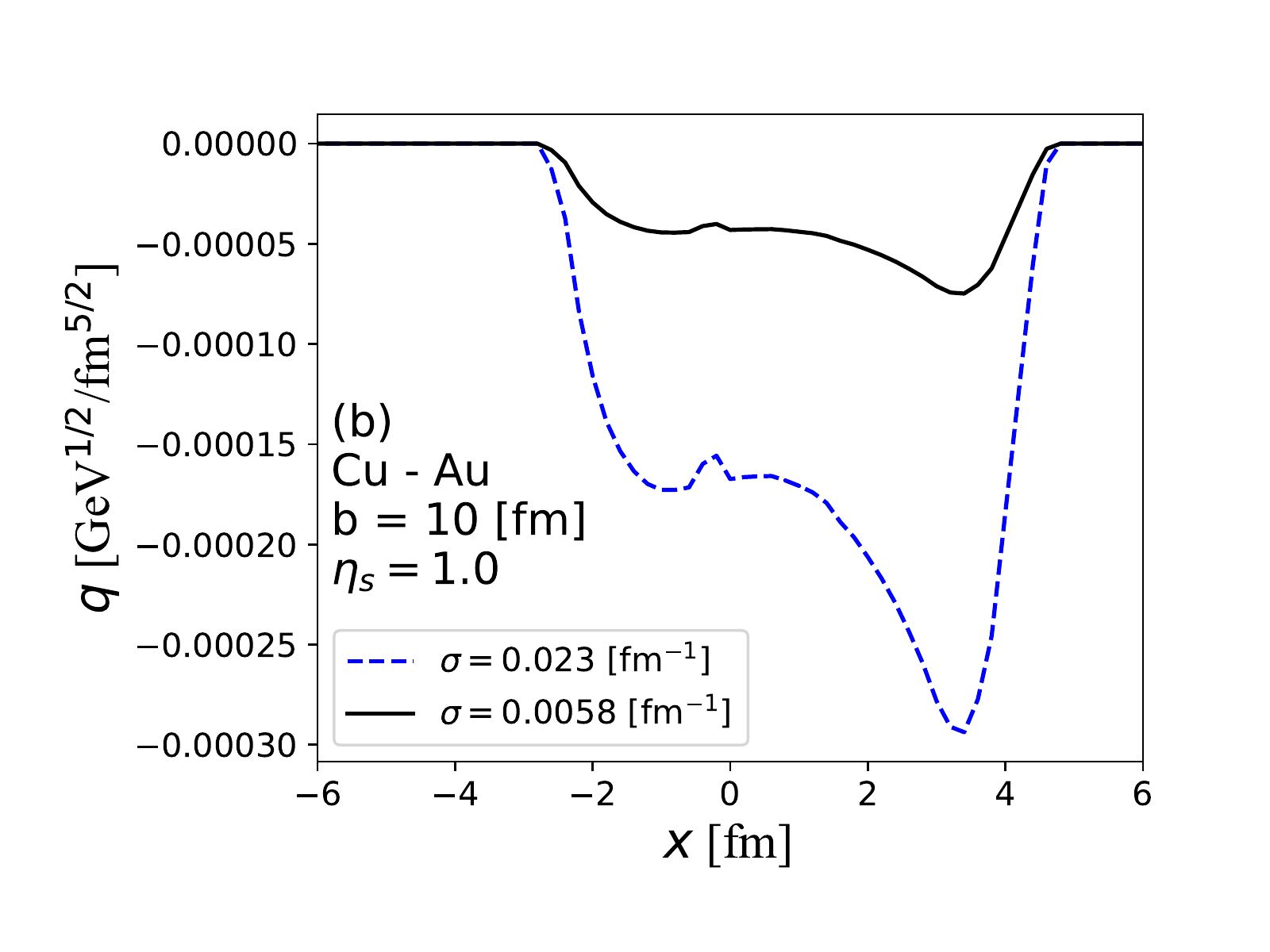}
     \end{minipage}
    \caption{(color online) The electric charge distributions on the freezeout hypersurface as a function of $x$ at $y = 0.0$ fm in the cases of (a) $\eta_s = -1.0$ and (b) $\eta_s = 1.0$ in Cu-Au collisions. The black solid and blue dashed lines represent $\sigma = 0.0058$ and $0.023~ \mathrm{fm}^{-1}$ cases.
    }
    \label{fig_hydro:q_freezeout surface eta10 Cu-Au}
\end{figure*}
Figures~\ref{fig_hydro:q_freezeout surface eta10 Cu-Au} (a) and (b) display the electric charge distribution on the freezeout hypersurface at $(y,\eta_s) = (0~\mathrm{fm}, -1)$ and $(0~\mathrm{fm}, +1)$ in Cu-Au collisions, respectively. 
In Cu-Au collisions, the electric charge density at $\eta_s = -1$ ($\eta_s = +1$) has only one local minimum in the negative (positive) $x$ region.
However, as shown in Fig.~\ref{fig_hydro:q_freezeout surface eta10 Cu-Au} (a), in $\eta_s < 0$, the slope of the electric charge profile is steeper than that of Au-Au collisions.
This reason is that as shown in Fig.~6 (b) of Ref.~\cite{Nakamura:2022idq}, in $\eta_s < 0$, the electric field in the positive $x$ side rapidly decreases with increasing $x$.
Besides, as shown in Fig.~\ref{fig_hydro:q_freezeout surface eta10 Cu-Au} (b), in $\eta_s > 0$, one can see the plateau in $x \in [-2,1.5]$.
The asymmetric profile of the electric field in Fig.~6 (b) of  Ref.~\cite{Nakamura:2022idq} is reflected to this structure.
The total absolute value of electric charge density at $\eta_s = 1.0$ in Fig.~\ref{fig_hydro:q_freezeout surface eta10 Cu-Au} (a) is larger than that at $\eta_s = -1.0$ in Fig.~\ref{fig_hydro:q_freezeout surface eta10 Cu-Au} (b). 
Since in Fig.~2 (b) of Ref.~\cite{Nakamura:2022idq}, the medium inside the freezeout hypersurface is close to the center of the Cu nucleus, the initial electric field produced by the Cu nucleus is dominated in $\eta_s > 0$.
It affects the electric density distributions.
The electrical conductivity dependence of electric charge distribution is the same as that in Au-Au collisions.

\subsection{The space-averaged velocity profile on the freezeout surface}
\begin{figure*}[t]
  \begin{minipage}[l]{0.45\linewidth}
    \includegraphics[width=8cm,height=6cm]{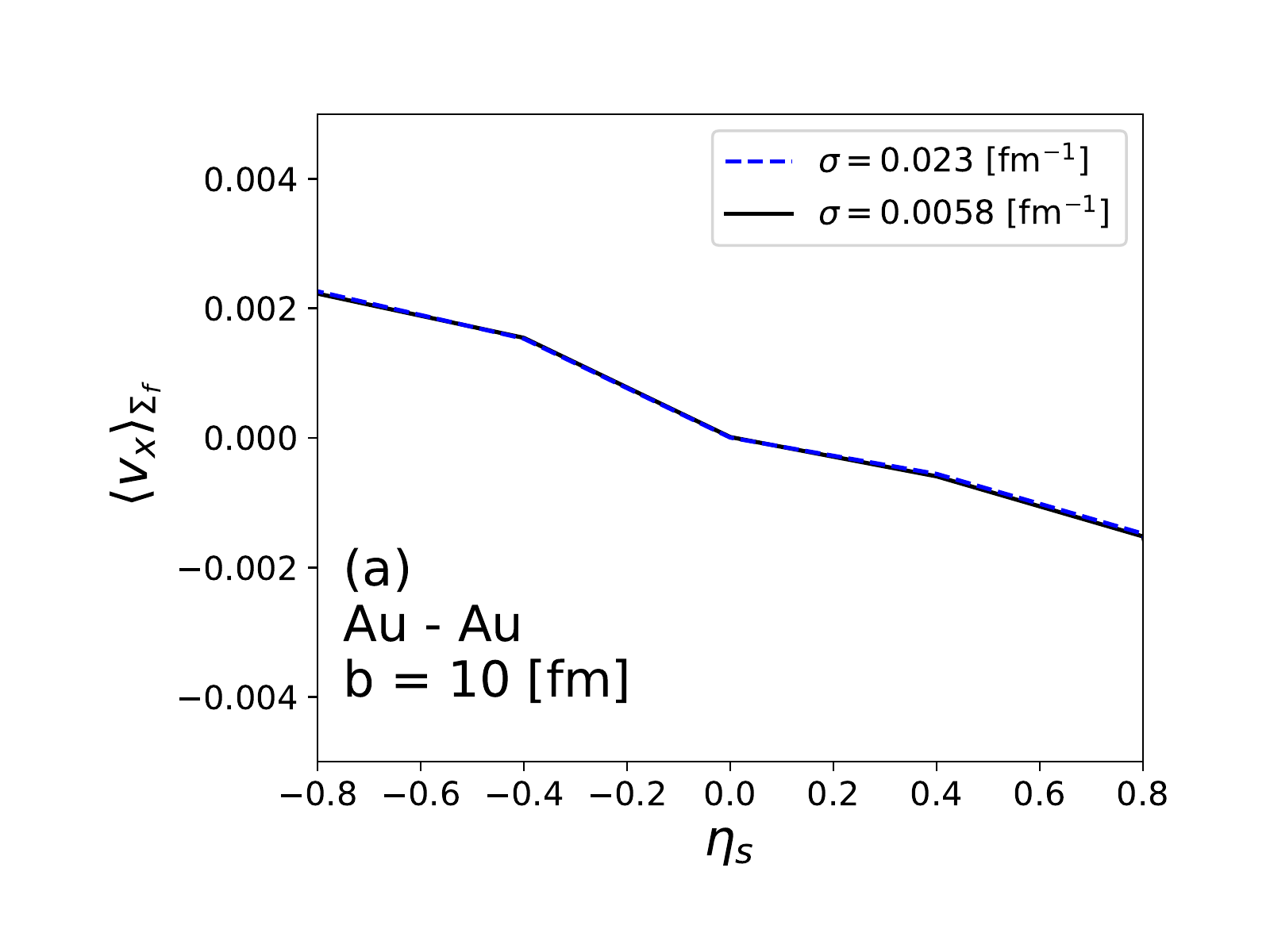}
  \end{minipage}
  \begin{minipage}[r]{0.45\linewidth}
    \includegraphics[width=8cm,height=6cm]{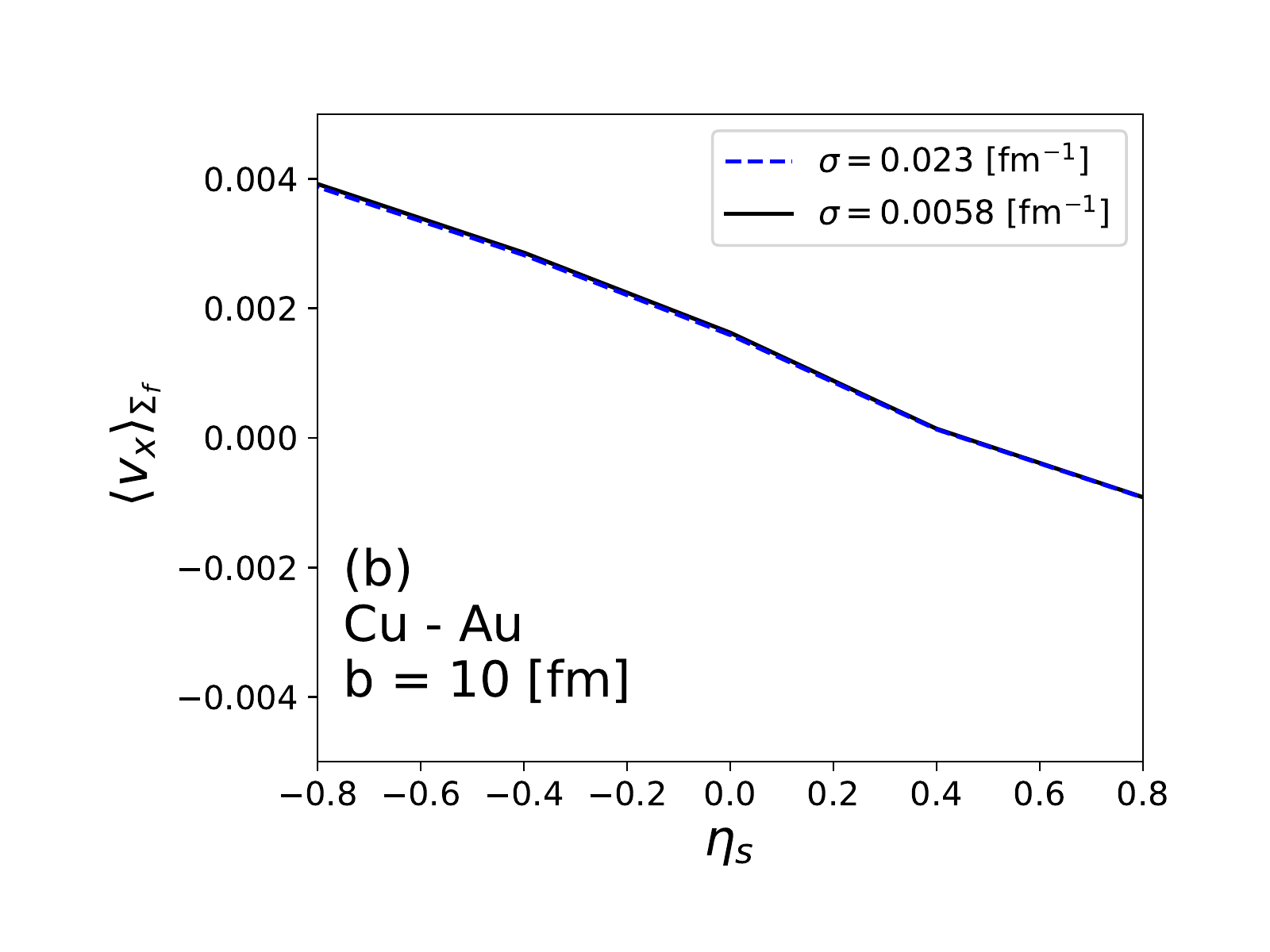}
     \end{minipage}
    \caption{(color online) The $x$-component of the space-averaged velocity on the freezeout hypersurface as a function of $\eta_s$ in (a) Au-Au and (b) Cu-Au collisions. The black solid and blue dashed lines represent velocity profiles in the cases of $\sigma = 0.0058$ and $0.023~ \mathrm{fm}^{-1}$.
    }
    \label{fig_hydro:vx_fs}
\end{figure*}
The electric charge density moves with the fluid velocity on the freezeout hypersurface, which becomes the source of the charge-dependent anisotropic flow.
We show the velocity profile on the freezeout hypersurface.
Figures~\ref{fig_hydro:vx_fs} (a) and (b) represent the profile of the space-averaged  velocity on the freezeout hypersurface in Au-Au and Cu-Au collisions, respectively. 
Here, we define the space-averaged velocity on the freezeout hypersurface as,
\begin{equation}
  \langle v_x \rangle_{\Sigma_f} = \frac{\int_{\Sigma_f} dydx \gamma e(x,y,\eta_s)v_x(x,y,\eta_s)}{\int_{\Sigma_f} dydx \gamma e(x,y,\eta_s)},
\end{equation} 
where $\Sigma_f$ is the freezeout hypersurface.
The initial tilted sources are reflected to the $x$-component of the space averaged velocity profiles in both of Au-Au and Cu-Au collisions.
In both cases, the electrical conductivity dependence is not observed.
In low electrical conductivity case, the velocity profile on the freezeout hypersurface is mainly determined by the pressure gradient of the QGP fluid.
The contribution of electromagnetic fields to the QGP fluid is evaluated by the plasma $\beta$, which is the ratio of the pressure to the energy density of electromagnetic fields, $\beta=p/p_{\mathrm{em}}$, where $p_{em}=\frac{\bm{E}^2 + \bm{B}^2}{2}$ is the energy density of electromagnetic fields.
In this calculations, since we take $e_0 = 55~\mathrm{GeV}/\mathrm{fm}^3$ and the magnetic field strength $|B^2|\sim 0.01~\mathrm{GeV}/\mathrm{fm}^3$, the plasma $\beta$ is evaluated to $\beta\sim 10^3$.
Hence, the contribution of electromagnetic fields becomes small.
Furthermore, in Au-Au collisions, the $x$-component of the space-averaged velocity has a negative value in $\eta_s > 0$ and a positive value in $\eta_s < 0$.
On the other hand, in $\eta_s < 0$, the negative charges are mainly produced in $x < 0$ fm region which is the opposite direction of the $x$-component of the space-averaged velocity. 
The magnitude of the directed flow of negatively charged hadrons has a smaller value than that of positively charged hadrons because of this structure between the electric charge distribution and the $x$-component of the space-averaged velocity profile.

In Au-Au collisions, the $x$-component of the space-averaged velocity is symmetric about $\eta_s = 0$. 
In Cu-Au collisions, the $x$-component of the space-averaged velocity has a positive value at $\eta_s = 0$ and vanishes near $\eta_s\sim 0.5$.
Since the initial asymmetric profile of the QGP medium on transverse plane provides the stronger pressure gradients with respect to the direction of the Cu side, the velocity is finite even at $\eta_s = 0$.

\subsection{Charge-dependent anisotropic flow}\label{Section: Charge-dependent anisotropic flow}
We investigate the effect of electromagnetic fields on the charge-dependent anisotropic flow.
The azimuthal anisotropic flow is calculated by,
\begin{equation}
    v_n(\eta) = \frac{\int dp_\mathrm{T}d\phi \cos(n\phi)\frac{dN}{dp_{\mathrm{T}}d\phi}}{\int dp_\mathrm{T}d\phi \frac{dN}{dp_{\mathrm{T}}d\phi}},
\end{equation}
where $p_\mathrm{T} =\sqrt{p_x^2+p_y^2}$, $\phi$, and $\eta = \frac{1}{2}\ln{\left(\frac{|\bm{p}|+p_z}{|\bm{p}|-p_z}\right)}$ are transverse momentum, an azimuthal angle with respect to the transverse plane, and the pseudorapidity of the hadrons, respectively.
To extract the purely magneto-hydrodynamic response, we ignore the final state interactions.
The hadron distribution is computed by using the information of the QGP fluid on the freezeout hypersurface in the Cooper-Frye prescription~\cite{Cooper:1974mv}.
Since the electric charge density is very small, $\sqrt{\hbar c}q/p\sim 10^{-4}$, we assume that the chemical potential of electric charge density is given by the linear relation to electric charge density, $\mu_q = q/g_\mathrm{h} T^2$, where $g_\mathrm{h}$ is the degree of freedom of hadrons and $T$ is the temperature of the medium.

\subsubsection{Charge-dependent elliptic flow}
\begin{figure}[h]
\includegraphics[width=8.5cm,height=6cm]{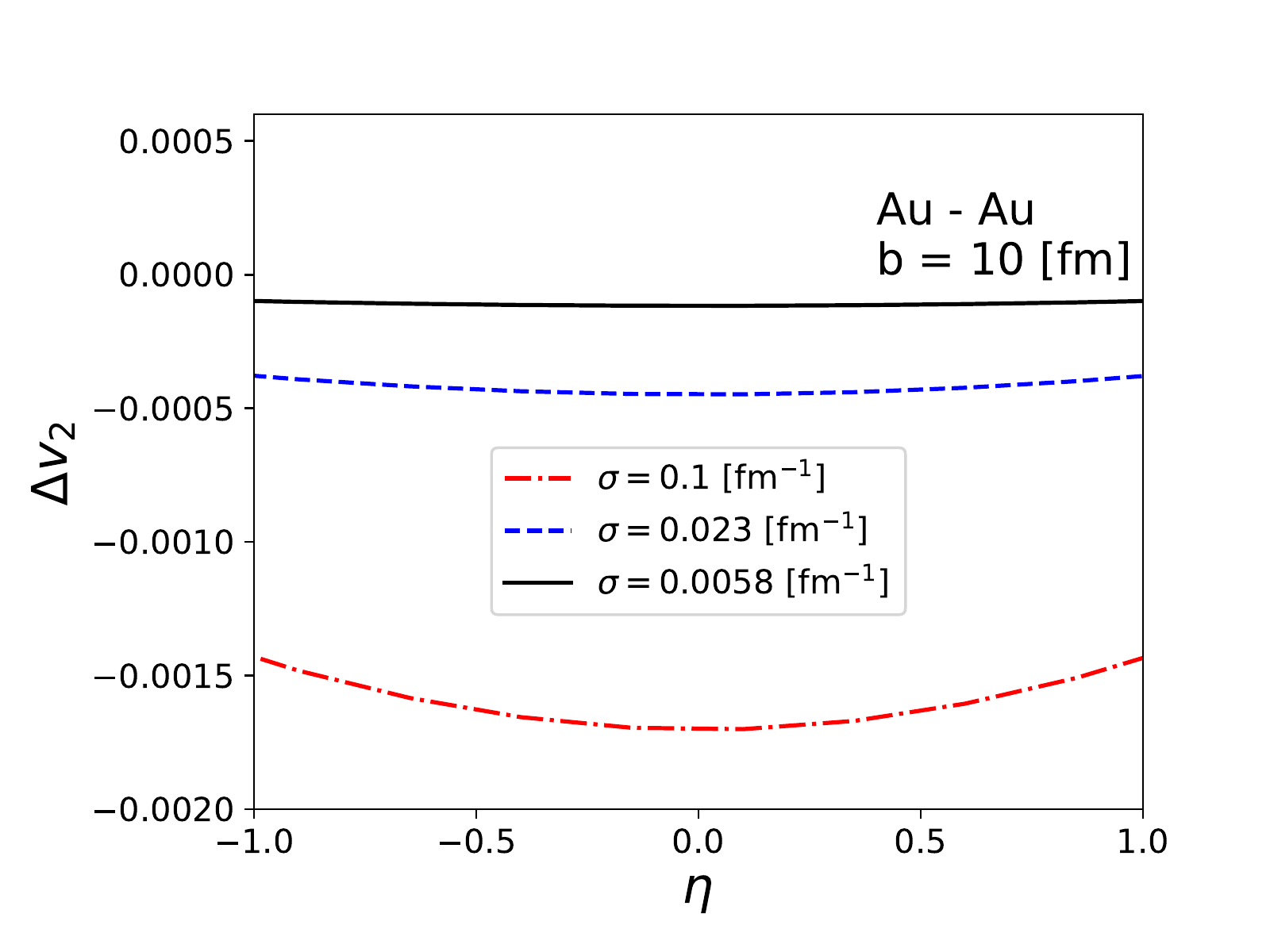}
\caption{\label{fig_hydro:charge-dep.elliptic flow Au+Au} (color online) The charge-odd contribution to the elliptic flow $\Delta v_2$ as a function of $\eta$ in Au-Au collisions. The black solid, blue dashed and red long dashed-dotted lines represent the charge-odd contribution to the elliptic flow in the cases of $\sigma = 0.0058$, $0.023$, and $0.1~\mathrm{fm}^{-1}$.}
\end{figure}
\begin{figure}[h]
\includegraphics[width=8.5cm,height=6cm]{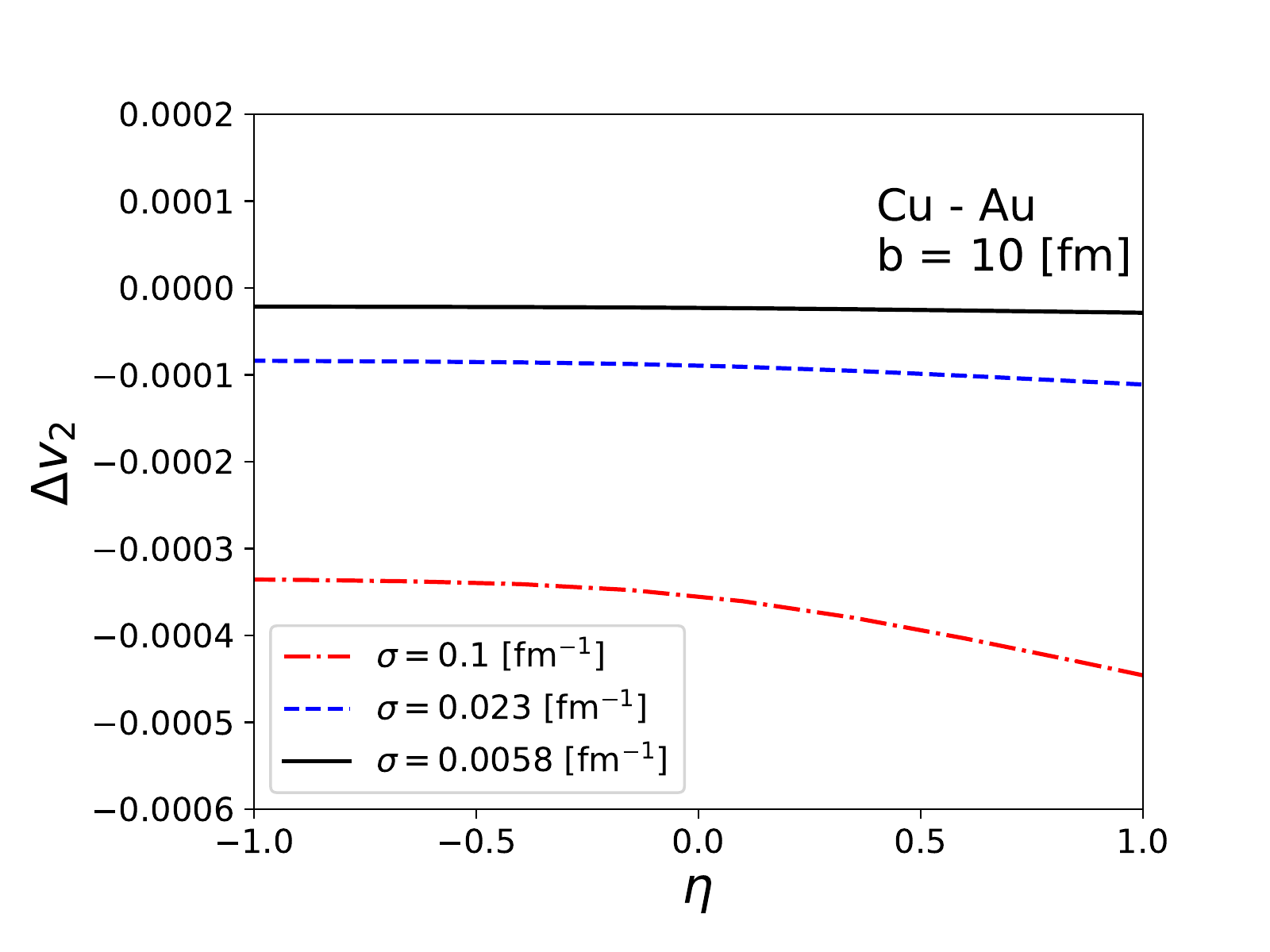}
\caption{\label{fig_hydro:charge-dep.elliptic flow Cu+Au} (color online) The charge-odd contribution to the elliptic flow $\Delta v_2$ as a function of $\eta$ in Cu-Au collisions. The black solid, blue dashed, and red long dashed-dotted lines represent the charge-odd contribution to the elliptic flow in the cases of $\sigma = 0.0058$, $0.023$, and $0.1~\mathrm{fm}^{-1}$.}
\end{figure}
Here, we consider the charge-odd contribution to the elliptic flow for $\pi$,
\begin{equation}
    \Delta v_2(\eta) = v_2^{\pi^+}(\eta) - v_2^{\pi^-}(\eta),
\end{equation}
focusing on the difference between the elliptic flow of $\pi^+$ and that of $\pi^-$. 
Figure~\ref{fig_hydro:charge-dep.elliptic flow Au+Au} represents the charge-odd contribution to the elliptic flow for $\pi$ as a function of $\eta$.
The black solid, blue dashed, red long-dashed dotted lines stand for the cases of $\sigma = 0.0058, 0.023$, and $0.1~\mathrm{fm}^{-1}$.
The electrical conductivity dependence is clearly observed.
The results in all electrical conductivity cases have negative values.
It implies that the elliptic flow of $\pi^-$ is enhanced by the conduction current associated with Ohm's law.
The behavior is also suggested by the electric charge distributions on the freezeout hypersurface in Fig.~\ref{fig_hydro:q_freezeout surface eta0}.
The value of the charge-odd contribution to the elliptic flow at $\eta = 0$ is consistent with the previous study in Ref.~\cite{Gursoy:2018yai}.
However, the rapidity dependence of the charge-odd contribution to the elliptic flow has the different tendency.
Our results show that the $|\Delta v_2 (\eta)|$ decreases with increasing $|\eta|$ but the result in Ref.~\cite{Gursoy:2018yai} show that the $|\Delta v_2 (\eta)|$ increases with $|\eta|$.
This is because the magnetic field increases with $|\eta_s|$ in Ref.~\cite{Gursoy:2018yai}. 
In our initial condition, electromagnetic fields decrease with increasing $|\eta_s|$ shown in Fig.~4 (b) of Ref.~\cite{Nakamura:2022idq}. 
Furthermore, as shown in Figs.~\ref{fig_hydro:q_freezeout surface eta10} (a) and (b), the total electric charge density is relatively small at $\eta_s = \pm1$ compared with that at $\eta_s = 0$ in Fig.~\ref{fig_hydro:q_freezeout surface eta0}.
Hence, the value of the charge-odd contribution to the elliptic flow at the finite $\eta_s$ becomes small. 

Figure~\ref{fig_hydro:charge-dep.elliptic flow Cu+Au} shows the charge-odd contribution to the elliptic flow for $\pi$ as a function of $\eta$ in Cu-Au collisions.
Clear electrical conductivity dependence is observed. 
As discussed in Figs.~\ref{fig_hydro:q_freezeout surface eta0 Cu-Au} and \ref{fig_hydro:q_freezeout surface eta10 Cu-Au}, the elliptic flow of the negatively charged hadrons is enhanced.
On the other hand, the production of positively charged hadrons reduces due to the negative chemical potential of electric charge density.
The elliptic flow of the positively charged hadrons decreases.
The charge-odd contribution to the elliptic flow of $\pi$ has a negative value.
In the forward rapidity region, the absolute value of the elliptic flow slightly increases in the cases of $\sigma = 0.023$ and $0.1~\mathrm{fm}^{-1}$.
This reason is that, in $\eta_s > 0$, the distribution of electric charge density has a plateau structure due to the electric field produced by the Cu nucleus in Fig.~\ref{fig_hydro:q_freezeout surface eta10 Cu-Au} (b).
This plateau makes the emission of negatively charged hadrons in direction of an $x$-axis negative direction increase.
Then, the elliptic momentum anisotropy of negatively charged hadrons becomes larger than that in $\eta_s < 0$ and at $\eta_s = 0$.
In both collisions, the charge-odd contribution to the elliptic flow is approximately proportional to the electrical conductivity.
The charge-odd contribution to the elliptic flow is sensitive to electrical conductivity.

\subsubsection{Charge-dependent directed flow}
\begin{figure}[ht]
\includegraphics[width=8.5cm,height=6cm]{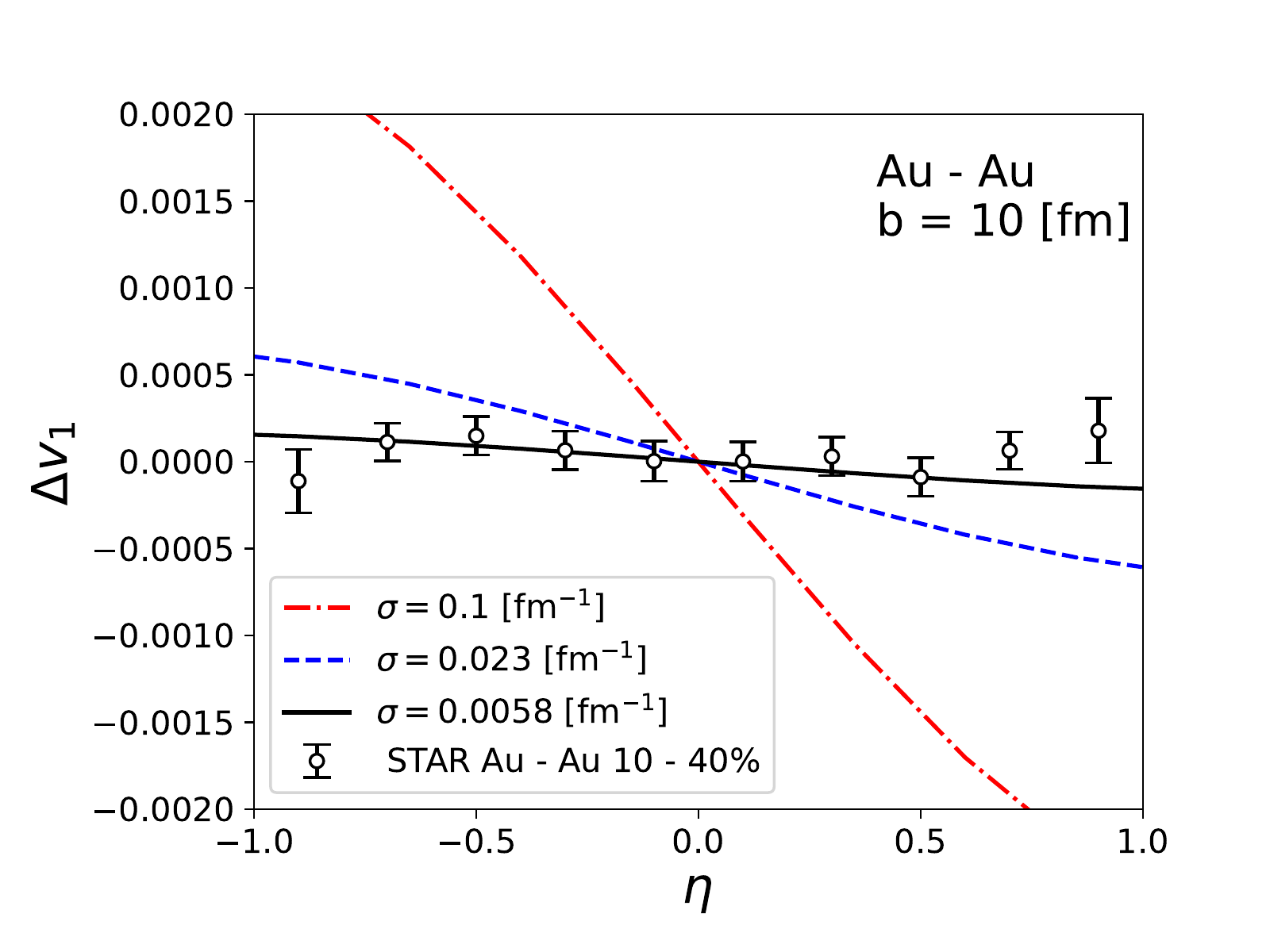}
\caption{\label{fig_hydro:charge-dep.directed flow Au+Au} (color online) The charge-odd contribution to the directed flow $\Delta v_1$ as a function of $\eta$ in Au-Au collisions. The black solid, blue dashed and red long dashed-dotted lines represent the charge-odd contribution to the directed flow in the cases of $\sigma = 0.0058$, $0.023$, and $0.1~\mathrm{fm}^{-1}$.}
\end{figure}
\begin{figure}[ht]
\includegraphics[width=8.5cm,height=6cm]{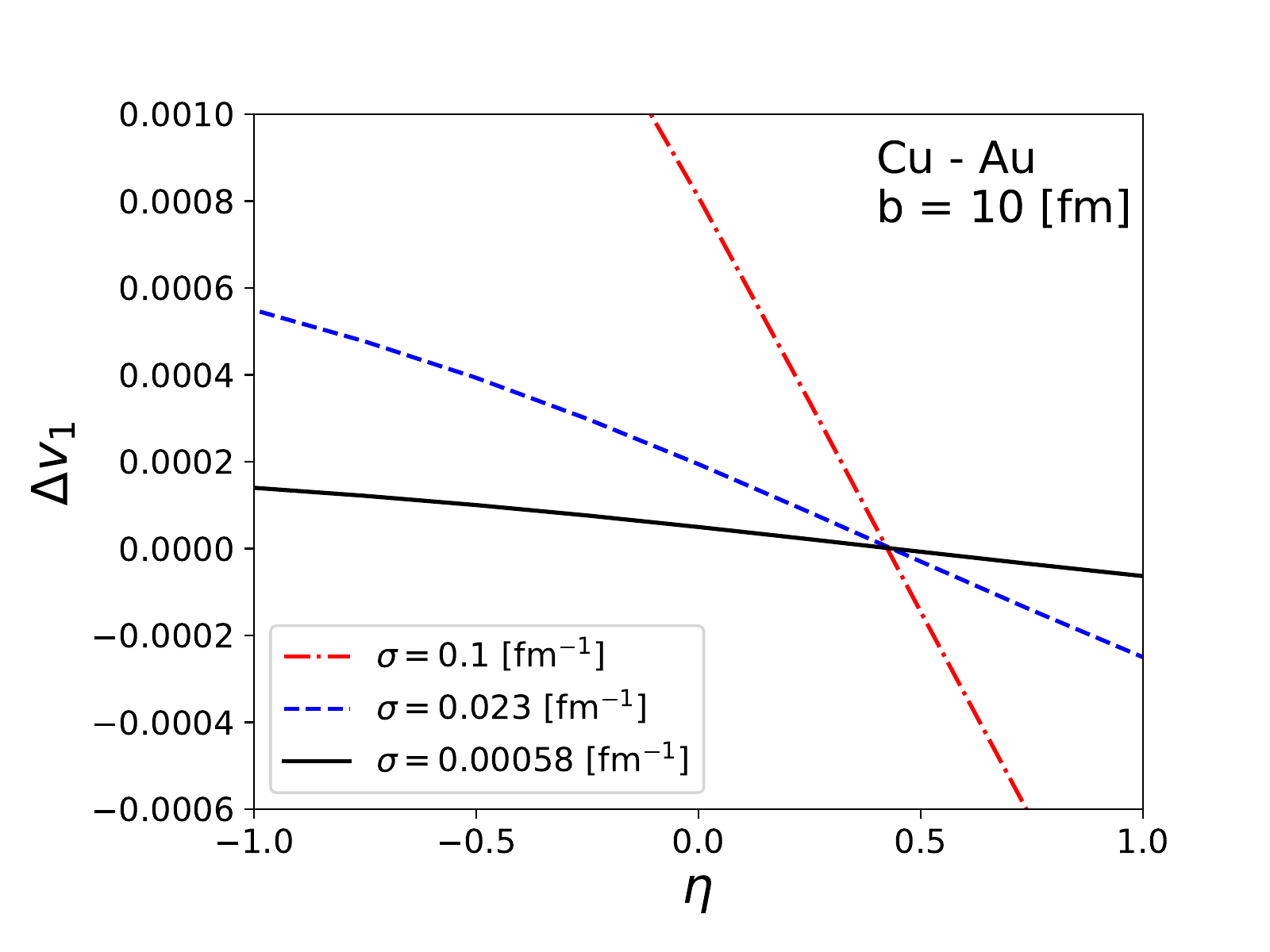}
\caption{\label{fig_hydro:charge-dep.directed flow Cu+Au} (color online) The charge-odd contribution to the directed flow $\Delta v_1$ as a function of $\eta$ in Cu-Au collisions. The black solid, blue dashed, and red long dashed-dotted lines represent the charge-odd contribution to the directed flow in the cases of $\sigma = 0.0058$, $0.023$, and $0.1~\mathrm{fm}^{-1}$.}
\end{figure}
Next we show the charge-odd contribution to the directed flow for $\pi$,
\begin{equation}
    \Delta v_1(\eta) = v_1^{\pi^+}(\eta) - v_1^{\pi^-}(\eta).
\end{equation}
Figure~\ref{fig_hydro:charge-dep.directed flow Au+Au} displays the charge-odd contribution to the directed flow of $\pi$ in Au-Au collisions.
The black solid, blue dashed, and red long-dashed dotted lines stand for the cases of $\sigma = 0.0058, 0.023$, and $0.1~\mathrm{fm}^{-1}$.
The tendency of the charge-odd contribution to the directed flow is similar to the charge-even contribution to the directed flow for $\pi$ shown in Fig.~10 (a) of Ref.~\cite{Nakamura:2022idq},
\begin{equation}
    v_1(\eta) = v_1^{\pi^+}(\eta) + v_1^{\pi^-}(\eta).
\end{equation}
This behavior is explained by the electric charge distribution on the freezeout hypersurface as discussed in Subsection~\ref{charge distribution on the freezeout surface}.
In Fig.~\ref{fig_hydro:q_freezeout surface eta0}, at $\eta_s = 0$, the electric charge density is symmetric about $x = 0$ fm.
For this reason, $\Delta v_1$ is zero at $\eta = 0$.
In the forward rapidity region, as shown in Fig.~\ref{fig_hydro:q_freezeout surface eta10} (b), the $\pi^-$ is emitted mainly in the positive $x$ region, though the $\langle v_x \rangle_{\Sigma_{f}}$ has a negative value in the forward rapidity region.
Hence, the contribution of the absolute value of the directed flow of the $\pi^-$ is slightly smaller than that of the $\pi^+$.
As a result, the charge-odd contribution to the directed flow has a negative value in the forward rapidity region.
On the other hand, in the backward rapidity region, the $\pi^-$ is produced mainly in the negative $x$ region.  
The $x$-component of space-averaged velocity has a positive value in the backward rapidity region. 
Then, the directed flow of the $\pi^-$ has a slightly smaller value than that of the $\pi^+$.
The charge-odd contribution to the directed flow becomes positive in the backward rapidity region.

The electrical conductivity dependence is clearly observed in the forward and backward rapidity regions.
The charge-odd contribution to the directed flow is approximately proportional to electrical conductivity.
This dependence is consistent with the previous study in Ref.~\cite{Gursoy:2018yai}.
Furthermore, we compare our results with the STAR data~\cite{PhysRevLett.112.162301}.
The result in the case of $\sigma = 0.023~\mathrm{fm}^{-1}$, which corresponds to $\sigma = (5.8 \pm 2.9)/\hbar c~\mathrm{fm}^{-1}$ of the three-flavor QGP at $T = 250~\mathrm{MeV}$ in the lattice QCD calculations~\cite{Aarts:2007wj,Ding:2010ga,Brandt:2012jc,FRANCIS2012212}, is slightly larger than that of STAR data.
In the lower conductive medium with $\sigma = 0.0058~\mathrm{fm}^{-1}$, our result is consistent with the STAR data within the error bar.
It implies the possibility of the incomplete electromagnetic response of the QGP medium~\cite{Akamatsu:2011nr, PhysRevC.105.L041901, Dash:2022xkz}, though there is still ambiguity to the conclusive value of electrical conductivity 
in lattice QCD calculation. 
In our model, the relaxation time of the electric current is ignored.
If the relaxation process of the electric current is included, the effective electrical conductivity becomes small, because of 
the suppression of electric current by the long-time electromagnetic response associated with the relaxation process. 
To discuss quantitatively this effect, we need to extend Ohm's law, including the relaxation time of the electric current.
In more precise measurements of charge-odd contribution to the directed flow, it can be detected in high-energy heavy-ion collisions.

Figure~\ref{fig_hydro:charge-dep.directed flow Cu+Au} shows the charge-odd contribution to the directed flow of $\pi$ in Cu-Au collisions.
The electrical conductivity dependence is clearly observed.
The charge-odd contribution to the directed flow of $\pi$ is approximately proportional to electrical conductivity at $\eta = 0$ in Cu-Au collisions.
It is consistent with the straightforward estimate of the charge-odd contribution of directed flow in Cu-Au collisions~\cite{Hirono:2012rt}.
The electric charge density at $\eta_s = 0$ shown in Fig.~\ref{fig_hydro:q_freezeout surface eta0 Cu-Au} is reflected to this electrical conductivity dependence.
The charge-odd contribution to the directed flow has the non-zero value at $\eta = 0 $ in finite electrical conductivity case.
There are two reasons.
First, one is that the $\langle v_x\rangle_{\Sigma_f}$ has the finite value at $\eta_s = 0$ in Fig.~\ref{fig_hydro:vx_fs} (b) due to the stronger pressure gradient along with the impact parameter on the side of the Cu nucleus in the initial energy density profile of the QGP medium.
Second, one is that the asymmetric structure of the electric charge distribution in Fig.~\ref{fig_hydro:q_freezeout surface eta0 Cu-Au} due to the initial electric fields at $\eta_s = 0$ is reflected to the $\Delta v_1$ at $\eta=0$. 
In all electrical conductivity cases, $\Delta v_1$ is crossing zero point near $\eta = 0.5$.
This reason is that the velocity on the freezeout hypersurface is vanishing near $\eta_s = 0.5$ in Fig.~\ref{fig_hydro:vx_fs} (b).
Hence, there is no electrical conductivity dependence since the velocity profile has no deviation in each electrical conductivity.
Furthermore, in Cu-Au collisions, the charge-odd contribution to the directed flow may be easily detected since it has the non-zero value at $\eta = 0$ in finite electrical conductivity case.
Our result indicates that the precise measurement of the charge-odd contribution to the directed flow is appropriate for the determination of the value of the electrical conductivity of the QGP. 
This measurement sheds light on the electromagnetic response of the QGP medium .

\section{Summary}\label{summary}
We have investigated the charge-dependent anisotropic flow, utilizing our RRMHD model for high-energy heavy-ion collisions~\cite{Nakamura:2022idq,Nakamura:2022wqr}.
In order to compare our results of the charge-odd contribution to the directed flow with STAR data, we focused on RHIC energy.
We considered the optical Glauber model~\cite{Glauber} as an initial condition of the QGP medium.
In order to study the directed flow, the tilted source was adopted in the longitudinal profile of the energy density~\cite{Bozek:2010bi}.
The solution of the Maxwell equations was taken to be an initial condition of electromagnetic fields~\cite{PhysRevC.88.024911}.
We considered the system in which the electric charge $q_0$ is moving along parallel to the beam axis ($\hat{\mathbf{z}}$) with velocity $v$ in the laboratory frame by an observer located at $\mathbf{r} = z\hat{\mathbf{z}} + \mathbf{x}_\perp$ in the Minkowski coordinates.
The parameters of the initial condition of the QGP medium have been determined from the comparison with the STAR data of the directed flow in Au-Au collisions~\cite{Nakamura:2022idq,STAR:2008jgm}.

The RRMHD simulation was performed with this initial condition in both of Au-Au and Cu-Au collisions.
The electrical conductivity was taken to be constant values, $\sigma = 0.0058, 0.023$ and $0.1~\mathrm{fm}^{-1}$.
We found that the electric charge distribution is sensitive to the RRMHD evolution and the initial conditions of the QGP medium and electromagnetic fields for the different collision systems.
The clear electrical conductivity dependence is observed in the electric charge distribution.
The electric charge distribution is approximately proportional to electrical conductivity.

We have calculated the charge-odd contribution to the anisotropic flows in Au-Au and Cu-Au collisions.
The charge-odd contribution to the directed flow and elliptic flow is sensitive to the electrical conductivity and the initial profile of electromagnetic fields for different collision systems.
We confirmed that the elliptic flow of the $\pi^-$ is enhanced by the conduction current associated with Ohm's law.
As a result, the charge-odd contribution to the elliptic flow has a negative value in both of Au-Au and Cu-Au collisions.
Besides, it is approximately proportional to electrical conductivity.
In the charge-odd contribution to the directed flow, the electrical conductivity dependence is also clearly observed in both collisions.
We compared our results with the STAR data in Au-Au collisions~\cite{STAR:2017ykf}.
The result in the case of $\sigma = 0.023~\mathrm{fm}^{-1}$ is slightly larger than that of the STAR data.
This value of the electrical conductivity corresponds to $\sigma = (5.8 \pm 2.9)/\hbar c~\mathrm{fm}^{-1}$ of the three-flavor QGP at $T = 250~\mathrm{MeV}$ in the lattice QCD calculations~\cite{Aarts:2007wj,Ding:2010ga,Brandt:2012jc,FRANCIS2012212}.
On the other hand, in the higher resistive case of $\sigma = 0.0058~\mathrm{fm}^{-1}$, our result is consistent with the STAR data within the error bar.
It implies that the incomplete electromagnetic response of the QGP medium~\cite{Akamatsu:2011nr, PhysRevC.105.L041901, Dash:2022xkz} appears in STAR data~\cite{STAR:2017ykf}.
In more precise measurements, the incomplete electromagnetic response can be detected in high-energy heavy-ion collisions.
We note that, in order to more precise quantitative analysis, we need to take into account the viscous effect and the final state interaction of hadrons after freezeout process.
In our calculation, we employ the ultrarelativistic ideal EoS.
The EoS based on lattice QCD simulations should be used to determine the value of the electrical conductivity~\cite{Borsanyi:2010cj,PhysRevD.85.054503,BLUHM2014157}.
We leave them for future works.
In Cu-Au collisions, we observed that the charge-odd contribution to the directed flow is approximately proportional to electrical conductivity at $\eta=0$.
It is consistent with the straightforward estimate of the charge-odd contribution of directed flow in Cu-Au collisions~\cite{Hirono:2012rt}.
We comment on the parameters of the initial condition in Cu-Au collisions~\cite{Nakamura:2022idq}.
To compare purely the magneto-hydrodynamic response in Cu-Au collisions with that in Au-Au collisions, we employ the same value of the parameters in Cu-Au collisions as that in Au-Au collisions.
We should adjust these parameters more precisely for the comparison with the STAR data.
Even though, our results show the strong impact of RRMHD in charge-odd contribution to elliptic and directed flows in both collision systems.
Then, we conclude that the charge-dependent anisotropic flow is a good probe to extract the electrical conductivity of QGP medium.



\section*{Acknowledgement}
The work of K.N. was supported in part by JSPS Grant-in-Aid for JSPS Research Fellow No.~JP21J13651.
This work was also supported by JSPS KAKENHI Grant Numbers, JP19H01928, JP20K11851 (T.M.), 
JP20H00156, JP20H11581, JP17K05438 (C.N.), JP20H00156, JP20H01941, JP20K11851, and JP21H04488 (H.R.T.).

\bibliography{charge-dep_flow}

\end{document}